# DIRAC-SCHRÖDINGER TRANSFORMATIONS IN CONTACTED GRAPHENE STRUCTURES

**Daniela Dragoman***

Univ. Bucharest, Physics Dept., P.O. Box MG-11, 077125 Bucharest, Romania

**Abstract**

At an interface between contacts and graphene, the mathematical equation that governs the propagation of electrons transforms from the Schrödinger to the Dirac equation. The condition of current probability conservation at such an interface does not determine uniquely the boundary conditions for the quantum wavefunction. We discuss the possible form of boundary conditions, determine its influence on the transmission coefficient of a contacted graphene structure and suggest that optical experiments on photonic crystals with Dirac points can help identifying, under certain circumstances, the proper boundary condition at graphene/electrode interfaces.

___

* email address: danieladragoman@yahoo.com

**Introduction**

Graphene is a unique material, in which the charge carriers obey a massless Dirac-like equation, the two spinor components of the wavefunction referring to the two triangular sublattices of the honeycomb crystal [1-2]. As a consequence, specific phenomena such as the Klein paradox or the anomalous Hall effect are encountered in this material. However, when contacted, the charge carriers entering the graphene sheet are Schrödinger-type electrons. The successive Schrödinger-Dirac and then Dirac-Schrödinger mathematical transformations of the equation governing the propagation of charge carriers from one contact to graphene and from graphene to the other contact, respectively, have as results the decrease of the transmission coefficient at values lower than 1 even at normal incidence and the reduction of the frequency cutoff of graphene devices [3-4]. A similar conclusion has been drawn regarding light propagation through photonic crystals with Dirac point [5]. Nevertheless, in the last case it was suggested that the boundary conditions, determined solely by the requirement of current conservation, are not unique.

Because of the practical implications of the problem of proper boundary conditions in graphene devices, we analyze in this paper the Schrödinger-Dirac transformation in contacted graphene structures. We consider time-independent ballistic propagation and unveil the consequences of different boundary conditions on the transmission coefficient dependence on the incident angle of electrons and on the effective masses and electron energy in the contacts. The current probability conservation requirement cannot determine by itself the correct boundary conditions and there is no physical argument to support one boundary condition among the many possible choices. Therefore, similarly to the exact analogies between optical beams and Schrödinger-type electrons [6], which led to the development of entire fields of research, a photonic analog of contacted graphene devices can shed some light on the proper boundary conditions at the interfaces between graphene and electrodes.

**Boundary conditions of charge carriers at graphene/electrode interfaces**

Let us consider a graphene sheet extending in the (*xy*) plane and surrounded by two metallic contacts, the graphene-electrode interfaces being located at $x = 0$ and $x = L$. All parameters referring to the left electrode, graphene, and right electrode will be labeled by subscripts 1, 2 and 3, respectively. In particular, the ballistic charge carriers in graphene propagating with energy $E$ satisfy the time-independent quantum Dirac equation

$$-i\hbar v_F \begin{pmatrix} 0 & \partial/\partial x - i\partial/\partial y \\ \partial/\partial x + i\partial/\partial y & 0 \end{pmatrix} \begin{pmatrix} \psi_{2+} \\ \psi_{2-} \end{pmatrix} = -i\hbar v_F (\nabla \cdot \boldsymbol{\sigma}) \begin{pmatrix} \psi_{2+} \\ \psi_{2-} \end{pmatrix} = (E - V_2) \begin{pmatrix} \psi_{2+} \\ \psi_{2-} \end{pmatrix}, \quad (1)$$

where $\psi_{2+}$ and $\psi_{2-}$ are the components of the spinorial wavefunction $\Psi_2 = (\psi_{2+}, \psi_{2-})^T$ (T denotes the operation of transposition), $\boldsymbol{\sigma} = (\sigma_x, \sigma_y)$ is the vector of Pauli matrices, and $v_F \cong c/300$ is the Fermi velocity of charge carriers in graphene. If the potential energy is denoted by $V_2$, the dispersion relation for plane-wave solutions of (1) is $E - V_2 = \pm \hbar v_F \sqrt{k_y^2 + k_2^2}$, where $k_2 = \text{sgn}(E - V_2)\sqrt{(E - V_2)^2/(\hbar v_F)^2 - k_y^2}$ [7] is the *x* component of the wavevector in graphene and $k_y$ is the *y* component of the wavevector, which remains constant throughout the structure. The charge carriers are electrons if $\text{sgn}(E - V_2) = 1$ and holes, otherwise. The probability current in the *x* direction (multiplied with the Fermi velocity for dimensionality reasons, as will become clear later on) is defined as $j_2 = v_F (\Psi_2^*)^T \sigma_x \Psi_2$, and takes the value $j_2 = v_F \cos\phi_2 (|A_2|^2 - |B_2|^2)$ for

$$\Psi_2 = 2^{-1/2} \exp(ik_y y) \begin{pmatrix} A_2 \exp(ik_2 x) + B_2 \exp(ik_2 x) \\ A_2 \exp(ik_2 x + i\phi_2) - B_2 \exp(-ik_2 x - i\phi_2) \end{pmatrix}, \quad (2)$$

where $\phi_2 = \arctan(k_y / k_2)$. We can write also the probability current as

$$j_2 = (\Phi_2^*)^T \sigma_x \Phi_2 \qquad (3)$$

where $\Phi_2 = (\varphi_{2+}, \varphi_{2-})^T = (\psi_{2+}, v_F \psi_{2-})^T$.

On the other hand, in electrodes the charge carriers satisfy the time-independent Schrödinger equation

$$\left[-\frac{\hbar^2}{2m_j}\left(\frac{\partial^2}{\partial x^2} + \frac{\partial^2}{\partial y^2}\right) + V_j\right]\Psi_j = E\Psi_j, \quad j=1,3, \qquad (4)$$

where $m_j$ and $V_j$ are the electrode effective masses and potential energies in electrode $j$, the probability current along $x$ being defined as $j_j = (i\hbar/2m_j)[\Psi_j(\partial \Psi_j^*/\partial x) - \Psi_j^*(\partial \Psi_j/\partial x)]$. For plane-wave solutions of the form $\Psi_j = \exp(ik_y y)[A_j \exp(ik_j x) + B_j \exp(-ik_j x)]$, with $k_j = \cos\phi_j \sqrt{2m_j(E-V_j)}/\hbar$ and $\phi_j$ the propagation angle in region $j$, the probability current in electrode $j$ is $j_j = v_j(|A_j|^2 - |B_j|^2)$, with $v_j = \hbar k_j / m_j$, $j = 1,3$; in the transmission region $B_3 = 0$. In particular, $j_j$ can be written as

$$j_j = (\Phi_j^*)^T \sigma_x \Phi_j, \qquad (5)$$

with

$$\Phi_j = 2^{-1/2} \exp(ik_y y) \begin{pmatrix} A_j \exp(ik_j x) + B_j \exp(-ik_j x) \\ v_j[A_j \exp(ik_j x) - B_j \exp(-ik_j x)] \end{pmatrix} = \begin{pmatrix} \varphi_{j+} \\ v_j \varphi_{j-} \end{pmatrix} \qquad (6)$$

The only requirement that must be satisfied, according to [5], is the conservation of the probability current at interfaces. In [3-4] it was shown that this condition can be satisfied if

$$\varphi_{1+}|_{x=0} = \psi_{2+}|_{x=0}, \quad v_1 \varphi_{1-}|_{x=0} = v_F \psi_{2-}|_{x=0}, \qquad (7a)$$

$$\psi_{2+}|_{x=L} = \varphi_{3+}|_{x=L}, \quad v_F \psi_{2-}|_{x=L} = v_3 \varphi_{3-}|_{x=L}. \qquad (7a)$$

In this case, the transmission probability is defined as $T = (v_3/v_1)|A_3/A_1|^2$ and is given by

$$T = \frac{4 v_1 v_3 v_F^2 \cos^2 \phi_2}{(v_1 v_3 + v_F^2)^2 \sin^2(k_2 L) + v_F^2 [v_1 \cos(k_2 L + \phi_2) + v_3 \cos(k_2 L - \phi_2)]^2} \qquad (8)$$

From this expression it follows that $T \leq 1$, the equality being obtained when $v_1 = v_F = v_3$, i.e. when velocity matching occurs throughout the structure. Velocity matching can be achieved for an appropriate Fermi energy level in contacts, which tunes the energy values of the incident electrons, or by using non-metallic electrodes, for example doped semiconductors or conducting polymers [8-9], to tune the effective electron mass to desired values.

The requirement of conservation of current probability at interfaces does not impose unique boundary conditions for the wavefunctions, and we have no other physical arguments to impose a certain boundary condition when a Dirac-Schrödinger transformation takes place. More precisely, the probability current is conserved whenever

$$\begin{pmatrix} \varphi_{1+} \\ v_1\varphi_{1+} \end{pmatrix}_{x=0} = \begin{pmatrix} \vartheta_{2+} \\ v_F\vartheta_{2-} \end{pmatrix}_{x=0}, \quad \begin{pmatrix} \vartheta_{2+} \\ v_F\vartheta_{2-} \end{pmatrix}_{x=L} = \begin{pmatrix} \varphi_{3+} \\ v_3\varphi_{3+} \end{pmatrix}_{x=L} \qquad (9)$$

where $\Theta_2 = (\vartheta_{2+}, \vartheta_{2-})^T = M\Psi_2$ with $M$ a unit-determinant matrix that satisfies the relation

$$M^{-1} = \sigma_x M^+ \sigma_x. \qquad (10)$$

From (10) it follows that the elements $M_{ij}$, $i,j = 1,2$ of the matrix $M$ obey the conditions: (i) $M_{11}$ and $M_{22}$ are real, (ii) $M_{12}$ and $M_{21}$ are imaginary.

With the boundary conditions (9) the transmission coefficient takes the form

$$T(M) = \frac{4v_1 v_3 v_F^2 \cos^2\phi_2}{|ia\sin(k_2 L) + [b\cos(k_2 L + \phi_2) - c\cos(k_2 L - \phi_2)]|^2}, \qquad (11)$$

where $a = v_F^2(M_{21}^2 - M_{22}^2) - v_1 v_3(M_{11}^2 - M_{12}^2) + v_F(v_1 - v_3)(M_{11}M_{21} - M_{12}M_{22})$, $b = v_F^2 M_{21}M_{22} - v_1 v_3 M_{11}M_{12} + v_F(v_1 M_{11}M_{22} - v_3 M_{12}M_{21})$, $c = v_F^2 M_{21}M_{22} - v_1 v_3 M_{11}M_{12} + v_F(v_1 M_{12}M_{21} - v_3 M_{11}M_{22})$. It can be easily checked that (8) is obtained for $M = I_2$, with $I_2$ the $2\times 2$ unit matrix, while a similar expression as in (8) but with $\phi_2$ replaced by $-\phi_2$ is retrieved for $M = i\sigma_x$, as mentioned previously. This last case corresponds to interchanging $\psi_{2+}$ and $\psi_{2-}$ in the boundary conditions (7), interchange that must be accompanied by a phase shift of $\pi/2$ at each graphene/electrode interface. The probability current is satisfied also for these new boundary conditions. This phase shift at graphene/contact interfaces related to the unitary condition of $M$, supplements the anomalous phase shifts already observed in graphene in connection with Berry phases in asymmetric structures [4], magnetic fields [10] or in the presence of dislocations [11].

Different possible boundary conditions in (9) lead to different dependences of $T$ on electron incidence angles and material parameters in the contacts. For instance, the contour plot of the transmission coefficient dependence on the incidence angle in the first electrode and the $M_{11}$ value for $M_{12} = M_{21} = 0$ is illustrated in Fig. 1(a), similar dependences on the imaginary part of $M_{12}$ for $M_{11} = 0$, $M_{22} = 1$ and on the imaginary part of $M_{21}$ for $M_{11} = 1$, $M_{12} = i$ being shown in Figs. 1(b) and 1(c), respectively. The angle $\phi_1$ is determined from the conservation of $k_y$ at propagation and in all cases the fourth element of the $M$ matrix is derived from the unitarity condition. The simulations were done for $V_1 = V_2 = V_3 = 0$, $E = 0.1$ eV, $L = 100$ nm, and $m_1 = m_3 = m_0$, with $m_0$ the free electron mass.

The plots in Fig. 1 have different shapes so that it is possible to identify the proper boundary conditions if it would be possible to fabricate ideal contacted ballistic devices with defect-free graphene flakes that make perfect contact to electrodes. Moreover, in order to decide upon the appropriate boundary condition, it would be necessary to inject and collect electrons propagating along different angles in the graphene sheet, which would imply sophisticated detecting devices and no scattering events throughout the structure. As discussed later on, a photonic analog could provide answers about the proper boundary conditions in contacted graphene flakes.

However, even if this analog photonic crystal slab can be fabricated, to distinguish between the situations $M = I_2$ and $M = i\sigma_x$ it is necessary to impose the condition $m_1 \neq m_3$, which leads to $v_1 \neq v_3$ (the contour plot showing the $T$ dependence on the incidence angle in the first electrode and the imaginary part of $M_{12}$ for $M_{11} = M_{22} = 0$ is identical to Fig. 1(a); see also equation (11)). This implies that the two electrodes are asymmetric. A contour plot of the transmission coefficient dependence on the incidence angle in the first electrode and the electron energy $E$ for $m_1 = m_0$ and $m_3 = m_0/2$ are shown in Figs. 2(a) for $M = I_2$, a very similar dependence being obtained for $M = i\sigma_x$. All other parameters are identical to those in the previous figure. The difference between the transmission coefficients for the $M = I_2$ and $M = i\sigma_x$ cases is represented by $\Delta T = T(M = I_2) - T(M = i\sigma_x)$, Fig. 2(b) illustrating its dependence on the same variables as in Fig. 2(a). Note that $\Delta T = 0$ at normal incidence, i.e. for $\phi_1 = 0$, and that $\Delta T$ varies between -0.1 and 0.1 in our simulation range. Although small, such $\Delta T$ values could be detected in a photonic analog of graphene.

Figure (2) shows also that the angular range of non-vanishing transmission coefficient widens as the electron energy increases. This range is limited by the critical angle for which $k_2$ becomes imaginary. Equivalently, the critical angle decreases as $m_1$ increases, as can be seen from Fig. 3 for $M = I_2$, $m_3 = m_0$ and the same values for other parameters as above.

Because the effective masses in contacts affect the transmission coefficient, the next question is for what values of these parameters $T$ is maximum? Evidently, high $T$ values are desirable in graphene-based devices. The dependence of the transmission coefficient on $m_1$ and $m_3$ has been plotted in Figs. 4(a) and 4(b) at normal incidence for $M = I_2$ (an identical dependence is obtained also for $M = i\sigma_x$) and for $M_{11} = 0$, $M_{12} = i$, $M_{21} = 0$, respectively. As mentioned previously, for $M = I_2$ a value of $T = 1$ irrespective of $L$ is obtained for velocity matching, when $v_1 = v_3 = v_F$, while a maximum transmission coefficient is observed in Fig. 4(b) for different effective masses in contacts, i.e. for an asymmetric structure. Again, measurements of $T$ at different angles and in different structures can decide on which of the boundary conditions occur in fact.

**Discussions and conclusions**

We have argued in the previous sections that in order to find the correct boundary conditions at interfaces between regions in which electrons satisfy Dirac and Schrödinger equations, experiments are necessary because the theory has no unique solution. Such experiments would require injection and detection of electrons at different angles in perfect graphene sheets contacted with electrodes with different effective masses. These requirements are extremely difficult to be met in contacted graphene structures, so that an optical analog would be helpful in order to establish the proper boundary conditions at propagation between regions in which photons are described by the Helmholtz and Dirac-like equations, respectively. In the optical case it is much easier to measure transmission coefficient at different angles than for electrons in solid-state devices. The required structure would be a photonic crystal with Dirac points, as that studied in [12-13], surrounded by homogeneous media with different dielectric constants. To obtain relevant information about ballistic electron propagation in graphene, the photonic crystal should incorporate magnetic materials and be excited with TE electromagnetic fields, or should consist of regions with different dielectric permittivities and be excited with TM radiation. The reason is that the boundary conditions in the outer regions (1 and 3) the effective mass variation present in the Schrödinger equation must be replaced with the variation of the respective dielectric constant [14]. Such structures have been in fact already studied. Dirac cones for TM excitation, for example, appear at the center of the first Brillouin zones in photonic crystals consisting of a square array of alumina cylinders [15] or at the edges of the first Brillouin zone in honeycomb photonic crystals with time-reversal symmetries [16-17].

Polarized light propagation experiments in photonic crystals with Dirac points surrounded by media with different dielectric constants could indicate the correct boundary conditions for charge carriers at graphene/electrodes interfaces. This assumption is supported by the well-established analogies between time-independent Maxwell and Schrödinger equations [6] as well as by the similar Dirac-like propagation laws in photonic crystals with Dirac points and graphene. The identification of the proper boundary conditions has important application in nanoelectronics and optoelectronics.

## Acknowledgements

This work was supported by a grant of the Romanian National Authority for Scientific research, CNCS-UEFISCDI, project number PN-II-ID-PCE-2011-3-0224

**Figure captions**

Fig. 1 Transmission coefficient dependence on the incidence angle and (a) the $M_{11}$ value for $M_{12} = M_{21} = 0$, (a) the imaginary part of $M_{12}$ for $M_{11}= 0$, and $M_{22}= 1$ (c) and the imaginary part of $M_{21}$ for $M_{11} =1$, $M_{12} =i$.

Fig. 2 Dependence on the incidence angle and the electron energy of (a) the transmission coefficient for $M = I_2$, $m_1 = m_0$ and $m_3 = m_0/2$ and (b) the difference between the transmission coefficients for $M = I_2$ and $M = i\sigma_x$.

Fig. 3 Transmission coefficient dependence on the incidence angle and $m_1$ increases for $M = I_2$ and $m_3 = m_0$.

Fig. 4 The dependence of the transmission coefficient on $m_1$ and $m_3$ at normal incidence for (a) $M = I_2$ and (b) $M_{11} = 0$, $M_{12} = i$, $M_{21} = 0$.

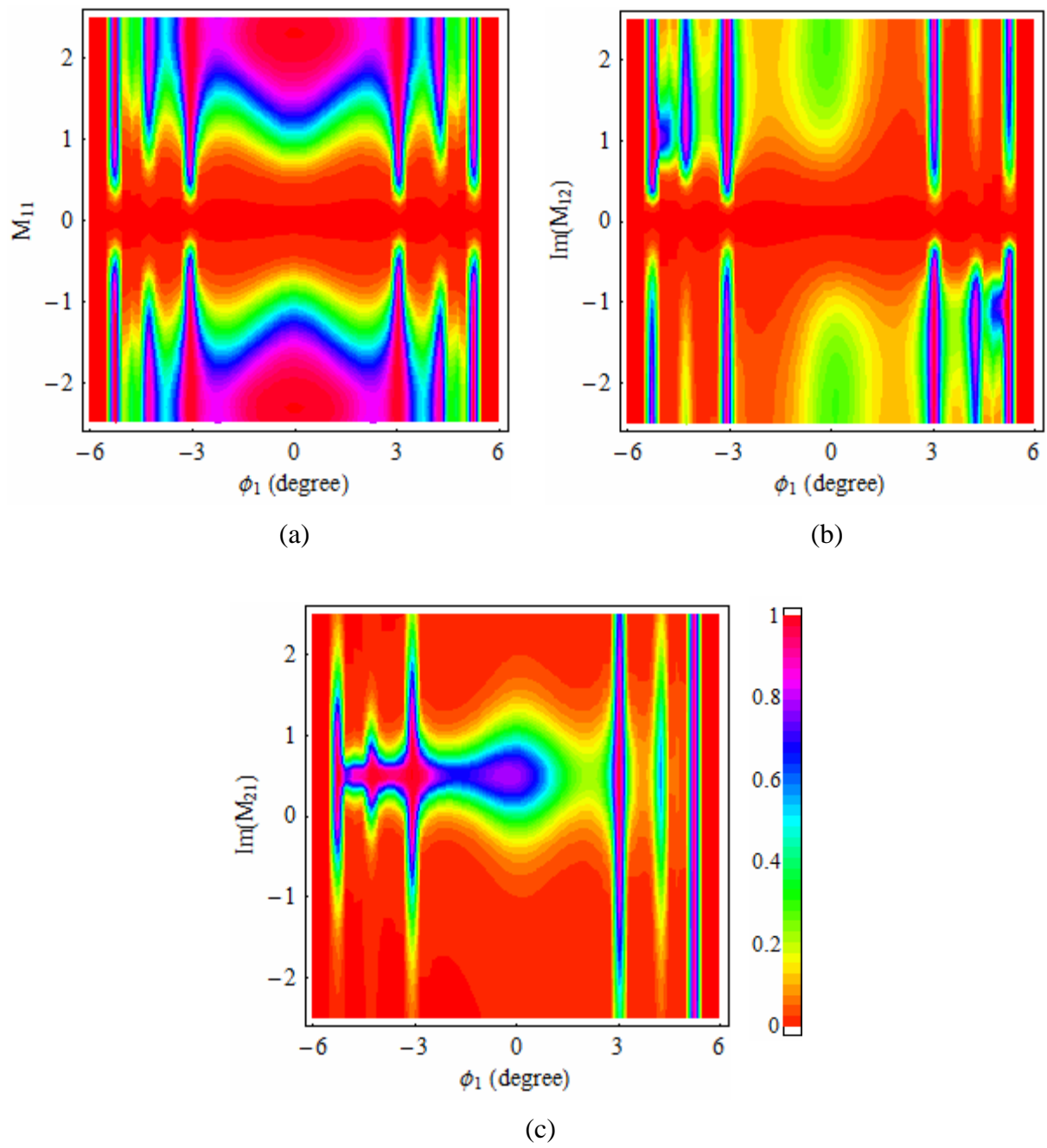

Fig .1

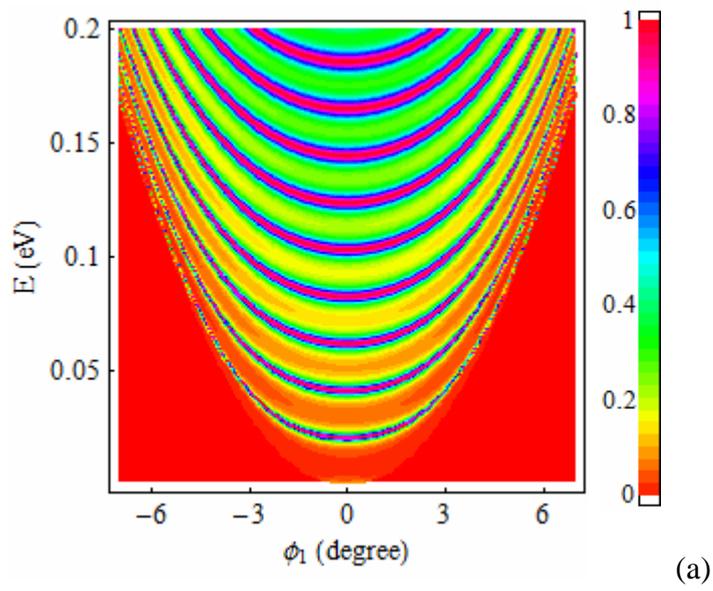

(a)

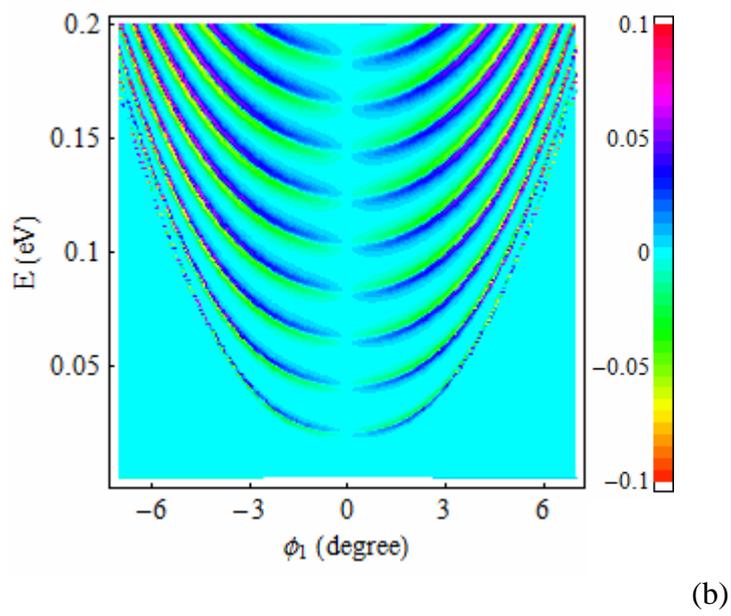

(b)

**Fig. 2**

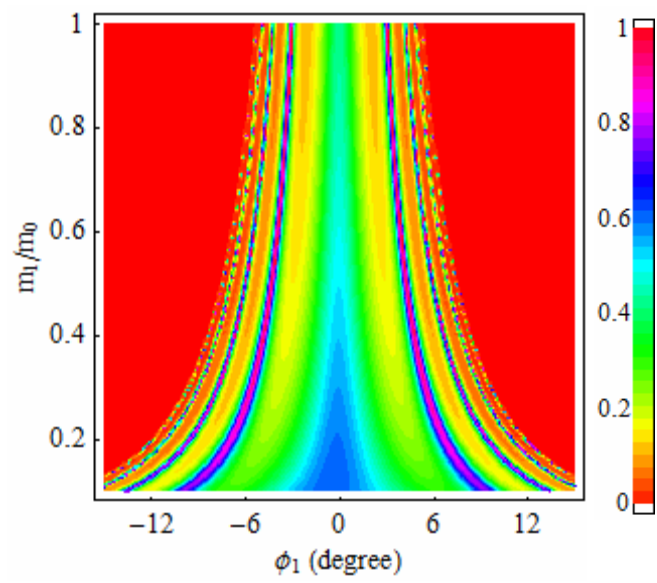

**Fig. 3**

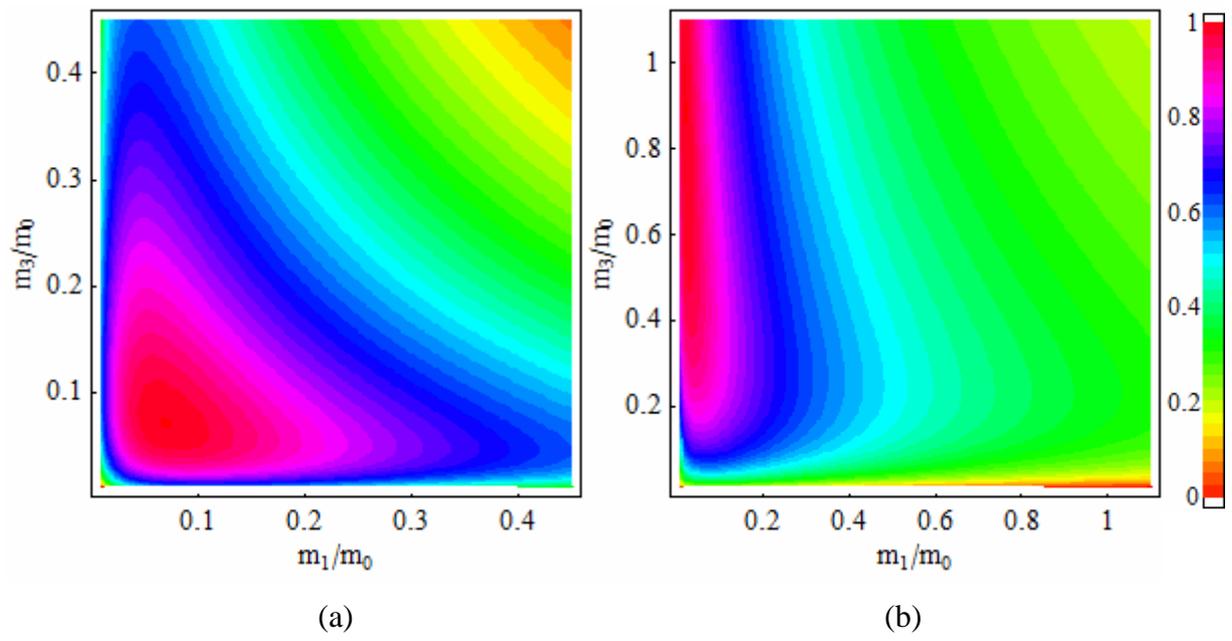

(a)                          (b)

**Fig. 4**